\documentclass[sigconf]{acmart}
\AtBeginDocument{%
  }

\copyrightyear{2025}
\acmYear{2025}
\setcopyright{rightsretained}
\acmConference[GECCO '25 Companion]{Genetic and Evolutionary Computation Conference}{July 14--18, 2025}{Malaga, Spain}
\acmBooktitle{Genetic and Evolutionary Computation Conference (GECCO '25 Companion), July 14--18, 2025, Malaga, Spain}
\acmDOI{10.1145/3712255.3726635}
\acmISBN{979-8-4007-1464-1/2025/07}

\begin{document}

\title{Finding Molecules with Specific Properties: Simulated Annealing vs. Evolution}


\author{Dominic Mashak}
\affiliation{%
  \institution{Southwestern University}
  \city{Georgetown}
  \state{Texas}
  \country{USA}}
\email{mashakd@southwestern.edu}

\author{S.A. Alexander}
\affiliation{%
  \institution{Southwestern University}
  \city{Georgetown}
  \state{Texas}
  \country{USA}}
\email{alexands@southwestern.edu}

\begin{abstract}
We compare the ability of a simulated annealing program and an evolutionary algorithm to find molecules with large molecular average hyperpolarizabilities. This property is an important component of nonlinear optical materials. Both optimization programs represent molecules as SMILES strings, a method that is widely used by chemists to describe molecular structure using short ASCII strings. Our results suggest that both approaches are comparable and can be used to solve a variety of more realistic problems of interest to chemists and material scientists.
\end{abstract}



\received{29 January 2025}
\received[revised]{1 April 2025}
\received[accepted]{4 April 2025}

\maketitle

\setlength{\tabcolsep}{3pt}

\section{Introduction}
Many quantum chemical programs can calculate the properties of a molecule from its structure but the reverse question, identifying a compound that has a specific set of properties, is a much more difficult problem because the solution space is vast and the variables are discrete rather than continuous. One way to approach this problem is to perform an elaborate trial-and-error search through an enormous collection of potential candidates. Another way is to use an optimization algorithm to sort through a set of potential molecules. These methods generally arrange a set of basic building blocks (atoms, chemical groups, etc.) using combinatorial methods \cite{Ref1,Ref2, Ref3}, simulated annealing \cite{Ref4,Ref5,Ref6} or genetic algorithms \cite{Ref7,Ref8,Ref9}.\par
Any optimization algorithm that wants to find molecules with specific properties must be able to represent the structure of a molecule in a form that a program can easily manipulate. SMILES (Simplified Molecular Input Line Entry Specification) is a simple language that describes the structure of chemical molecules using short ASCII strings. Four simple rules define a valid SMILES string \cite{Ref10}:
\begin{enumerate}
\item Atoms are described by their standard atomic symbol. Each symbol is normally enclosed in square brackets, such as [Au] for gold, however, many of the atoms found in organic molecules (such as B, C, N, O, P, S and F) are written without brackets. Most SMILES strings omit all hydrogen atoms. The implicit number of hydrogen atoms attached to other atoms is the difference between the atom's valence and the number of bonds assigned to the atom.
\item Single, double, and triple bonds are represented by the symbols `-', `=' and `\#' respectively. The atoms connected by these bonds are indicated by their adjacency. In most versions of SMILES, single bonds are omitted from the string, but for convenience, we explicitly show all bonds.
\item Branching is specified by placing the symbols for the atoms and bonds in this subchain between parentheses. These parentheses are placed directly after the symbol for the atom in the main sequence to which it is connected.
\item Rings are represented by breaking a single bond in each ring and then designating the two atoms connected by this bond with a digit immediately following the symbol for the atoms.
\end{enumerate}\par
In this paper, we compare the performance of simulated annealing with an evolutionary algorithm to find molecules with large hyperpolarizabilities, $\beta$. This property determines how a molecule interacts with light; molecules with large values can dramatically modify the frequency, phase and/or polarization of light. Several theoretical and experimental studies have identified numerous organic molecules that could form the basis of nonlinear optical (NLO) materials \cite{Ref11,Ref12}. Although these materials have several physical requirements (such as high thermal stability and transparency), a large hyperpolarizability is a critical one. For this reason, we want to determine which program increases $\beta$ as quickly as possible. For testing purposes, we consider molecules that contain only carbon, oxygen and hydrogen atoms. In Section 2 we examine a basic evolutionary algorithm and in Section 3 a simulated annealing program. In Section 4 we compare the results of both approaches.

\section{Evolutionary Algorithm Results}
Evolutionary algorithms were introduced in 1975 by John Holland \cite{Ref13}. In his book ``Adaptation in Natural and Artificial Systems'', Holland described how simulating biological evolution can become a general problem-solving strategy. Using these ideas, our program consists of the following basic steps:
\begin{enumerate}
\item Choose an initial population of parents. In our calculations, each parent is described as a SMILES string.\par\noindent
\item Apply mutation and crossover operators on the parents to generate a population of children. The choice of mutation or crossover is determined by a random number. If this number is below some fixed ratio, the seven mutation operators described below act on the parent:
\begin{itemize}
\item Pick a random bond and change it into a different type of bond (e.g. C=C-N-O $\rightarrow$ C-C-N-O)
\item Add a random atom with a random bond to a random location in the SMILES string\par (e.g. C=C-N-O $\rightarrow$ C=C-O-N-O)
\item Add a random atom with a random bond as a new branch to a random location in the SMILES string\par (e.g. C=C-N-O $\rightarrow$ C=C-N(-O)-O)
\item Delete a random atom and its connecting bond\par (e.g. C=C-N-O $\rightarrow$ C=C-N)
\item Pick a random atom in the SMILES string and change it into a different type of atom\par (e.g. C=C-N-O $\rightarrow$ C=C-N-C)
\item Pick two random atoms in the SMILES string and connect them with a ring (e.g. C=C-N-O $\rightarrow$ C1=C-N-O1)
\item Delete a ring (e.g. C1=C-N-O1 $\rightarrow$ C=C-N-C)
\end{itemize}
All valid strings generated by these mutations are included as children. If the random number is above the fixed ratio, a basic “cut and splice” crossover operator \cite{Ref14} is employed. This method chooses two strings at random, A and B, and breaks each at a random point between a bond and an atom. This ensures that the two ends of each string have a good chance of forming a valid string. The pieces, (A1, A2) and (B1, B2), are then recombined into the two new individuals (A1, B2) and (B1, A2). All valid strings are included as children and all invalid strings are discarded.\noindent
\item Apply a selection process to the entire population of children. Those with high fitness functions survive; those with low scores are discarded. The survivors then become the parents of the next generation. In our program, we divide the children into four equal groups. From the group with the highest fitness values, we randomly select 40\% of the children to survive. From the second group, we randomly select 30\% of the children; from the third group 20\% of the children; and from the fourth group 10\% of the children. This process ensures that high-fitness individuals form the bulk of the next generation but do not dominate it. By allowing some less fit individuals to be present, our evolutionary program is able to maintain a diverse population and reduce the possibility of getting trapped in local minima \cite{Ref15,Ref16}.\noindent
\item Return to Step 2 until the fitness function converges or the program reaches the maximum number of generations.\noindent
\end{enumerate}
Our evolutionary algorithm contains 10 parents and 20 children and runs for 100 generations. The initial parents are listed in Table 1 and were chosen from molecules that three research groups have identified as promising NLO materials \cite{Ref17,Ref18,Ref19}. We then repeated this calculation with four other random number seeds so that we could study its convergence.\par
There are several quantum-chemical techniques that can calculate this property; in general methods that can determine $\beta$ to high accuracy require large amounts of CPU time to evaluate even a modest size molecule. For this reason, we calculate $\beta$ using MOPAC, an open-source semiempirical program, and the PM6 force field\cite{Ref20}. As a result, our hyperpolarizability values are in atomic units. The amount of time required by these MOPAC calculations is still considerable, which is why we limited the number of generations to 100 and the number of children to only 20.\par
The values obtained by this calculation are presented in Table 2 as a function of the number of generations. As Figure 1 shows, the mutation/crossover ratio has a noticeable impact on the convergence. Although the 10/90 ratio clearly produces the fastest increase in $\beta$, the rate of improvement is about the same for all ratios until about generation 55. After that, the program was able to find a SMILES string that could serve as a suitable template for rapid progress. One noticeable feature of the 10/90 calculations is their exceptionally large standard deviation. This is because the different random number seeds produced a wide range of final values (649417, 33592, 28174, 133930 and 461557) that yield an average increase of 63\% in $\beta$.  

\begin{table*}
  \caption{Initial molecules used in our evolutionary algorithm}
  \label{tab:commands}
  \begin{tabular}{cc}
  \toprule
    Initial Molecule& $\beta$ \\
  \midrule
    C-N(-C)-C1=C-C=C(-C=C1)-N=C2-C=C-C(=O)-C=C2& 3389.13\\
    C1=C-C(-C=C-C2-C=C-C(-C(-O)-O)-C=C2)=C-C=C1-N(-C)-C& 1107.95\\
    C1=C-C(-C=C-C=C-C=O)=C-C=C1-N(-C)-C& 4137.17\\
    C1=C-C(-N(-O)-O)=C-C=C1-N& 470.46\\
    C1=C-C(-N(-N))=C-C=C1-N(-O)-O& 693.03\\
    C1=C-C(-C=C(-C)-C)=C-C=C1-N(-C)-C& 567.03\\
    C-N(-C)-C1=C-C=C(-C=C1)-N=N-C2=C-C=C(-C=C2)-N(-O)-O& 3656.56\\
    N-C1=C-C=C(-C=C1)-C=C-C2=C-C=C(-C=C2)-N(-O)-O& 1479.04\\
    C1=C-C(-C=C-N(-O)-O)=C-C=C1-N(-C)-C& 1423.70\\
    C1=C-C(-C=C-C=C-N(-O)-O)=C-C=C1-N(-C)-C& 2300.20\\
  \bottomrule
  \end{tabular}
\end{table*}

\begin{table*}
  \caption{Evolutionary calculation of the average hyperpolarizability (and its standard deviation) for 100 generations. These values have been averaged over 5 different random number seeds. The N is the number of function evaluations.}
  \label{tab:commands}
  \begin{tabular}{lcccccc}
  \toprule
    Generations& 0& 20&  40&  60&  80&  100\\
  \midrule
    F(x) 10\%& 4137$\pm$0& 13136$\pm$2725& 38140$\pm$45848& 82607$\pm$86271& 165544$\pm$175763& 261334$\pm$279803\\
    F(x) 30\%& 4137$\pm$0& 12804$\pm$1922& 21728$\pm$8152& 25098$\pm$7584& 30183$\pm$8321& 97176$\pm$14502\\
    F(x) 50\%& 4137$\pm$0& 14239$\pm$7082& 20186$\pm$10561& 23483$\pm$8538& 33802$\pm$13554& 42460$\pm$20356\\
    F(x) 70\%& 4137$\pm$0& 10802$\pm$2770& 16606$\pm$2577& 19158$\pm$6774& 20569$\pm$7812& 53666$\pm$59090\\
    N& 10& 410& 810& 1210& 1610& 2010\\
  \bottomrule
  \end{tabular}
\end{table*}

\begin{figure}[h]
  \centering
  \includegraphics[width=\linewidth]{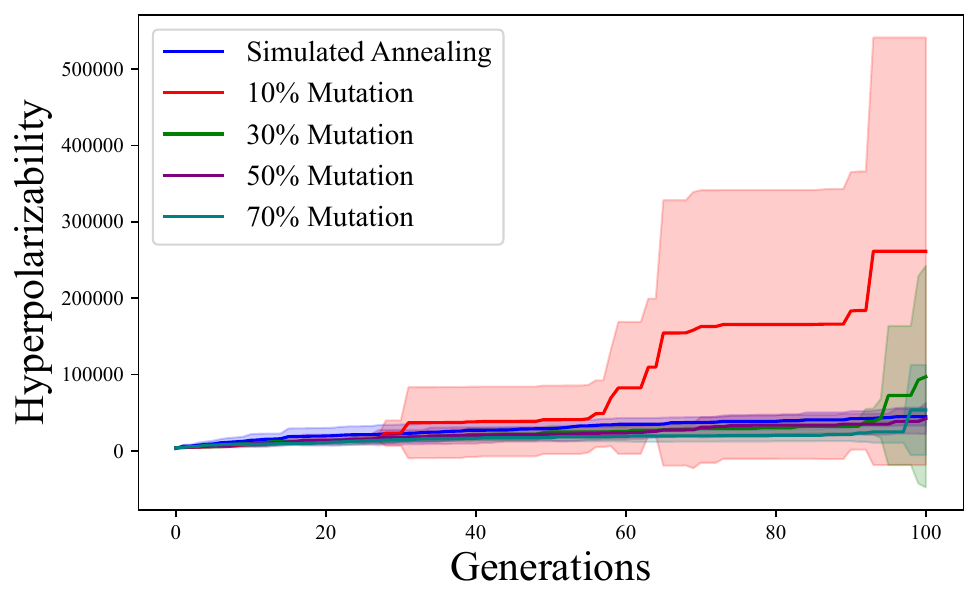}
  \caption{Average molecular hyperpolarizability (and its standard deviation) vs the number of generations. This illustrates the effect of the mutation/crossover ratio on the convergence. The values for this figure are summarized in Table 2 and have been averaged over 5 different random number seeds.}
\end{figure}

\section{Simulated Annealing Results}
Simulated annealing is a numerical optimization method that can find the global minimum of complicated multidimensional functions \cite{Ref21,Ref22}. Starting from some initial point, x$_{initial}$ and its value at that point, f(x$_{initial}$), this minimization method generates a new point in the multidimensional space, x$_{new}$, and calculates its value at that point, f(x$_{new}$). If f(x$_{initial}$) $>$ f(x$_{new}$), this step is accepted and x$_{new}$ becomes the starting point for the next step. If f(x$_{initial}$) $<$ f(x$_{new}$), an acceptance function determines whether x$_{new}$ is accepted or rejected. Simulated annealing tries to avoid getting stuck in a local minimum by occasionally accepting steps that yield worse solutions. In the acceptance function used by Metropolis et al. \cite{Ref23}
\[ A = min(R, exp([f(x_{new})-f(x_{initial})]/T)) \hspace{1cm} (1) \]
\par\noindent
T is a parameter known as the temperature and R is a random number between 0.0 and 1.0. If the value of R is greater than the exponential, the new point becomes the initial point in the next step, even though it has a higher fitness score. If R is less than the exponential function, the previous point is retained in the next step.\par
Most simulated annealing programs start at some high initial temperature. This normally allows a sequence of steps to effectively sample the entire parameter space, since most solutions are accepted. After a predetermined number of steps, the temperature is repeatedly reduced. In this paper we are only interested in improving the fitness function of each initial molecule, so we simply perform a short 100 step calculation at a fixed temperature of T=20. This temperature was previously shown to give good convergence \cite{Ref6}. During each step, we use the seven different mutation operators described in the previous section to convert the current SMILES string into a set of new SMILES strings.\par
For these calculations, we choose our fitness function to be the average hyperpolarizability, i.e. f(x$_{i}$)=$\beta$, and our initial molecule to be the first entry in Table 1. As before, we repeated this calculation with four other random number seeds so that we could study its convergence. Our results are presented in Table 3 and show that this method produces an average improvement of 13\%. Unlike the evolutionary results in the previous section, the different random number seeds produced a smaller range of final values (51041, 53993, 39607, 53337 and 26833) so the standard deviation is also smaller.

\begin{table*}
  \caption{Simulated annealing calculation of the average hyperpolarizability (and its standard deviation) as a function of the number of steps. These values have been averaged over 5 different random number seeds. N is the number of function evaluations.}
  \label{tab:commands}
  \begin{tabular}{lcccccc}
  \toprule
    Steps& 0& 20&  40&  60&  80& 100\\
  \midrule
    F(x)& 3389$\pm0$& 19920$\pm10481$& 27113$\pm11007$& 34907$\pm8136$& 39490$\pm8380$& 44962$\pm11683$\\
    N& 1$\pm0$& 136$\pm5$& 266$\pm6$& 393$\pm10$& 520$\pm9$& 641$\pm9$\\
  \bottomrule
  \end{tabular}
\end{table*}

\begin{figure}[h]
  \centering
  \includegraphics[width=\linewidth]{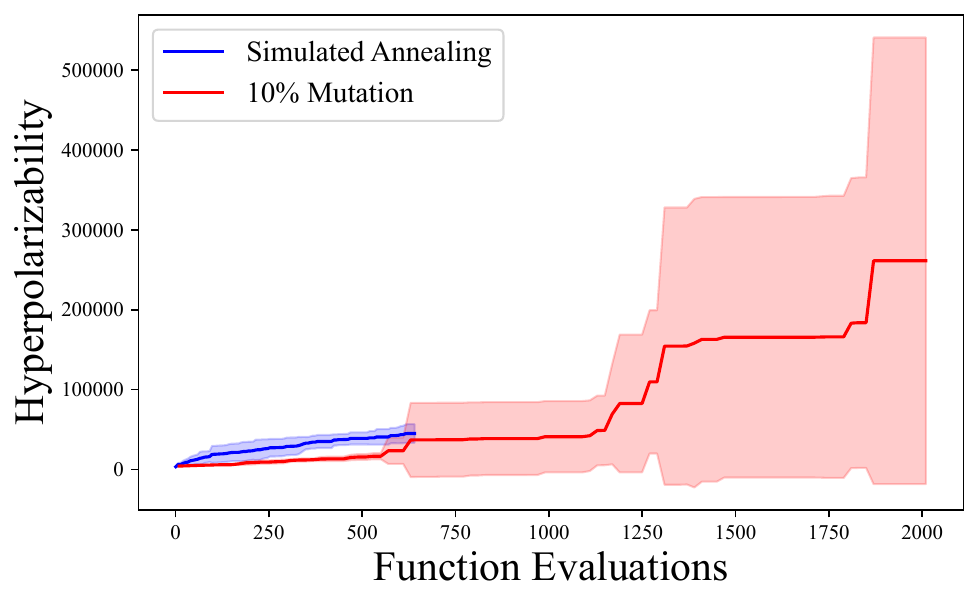}
  \caption{Average molecular hyperpolarizability (and its standard deviation) vs the number function evaluations. This illustrates the relative convergence of simulated annealing and the evolutionary algorithm from the perspective of the number of function evaluations. The values for this figure are summarized in Table 2 and 3 and have been averaged over 5 different random number seeds.}
\end{figure}

\section{Conclusions}
Since both simulated annealing and evolutionary algorithms make few assumptions about the function to be optimized, they are quite robust. They can also solve problems where the variables are discrete rather than continuous. These characteristics make them very suitable for designing molecules with specific properties. A cursory comparison between the values in Tables 2 and 3 shows that after 100 generations our evolutionary algorithm improves $\beta$ by 63\% and after 100 steps simulated annealing improves it by 13\%. If we plot $\beta$ with respect to the number of function evaluations, we see in Figure 2 that simulated annealing performs slightly better than the 10/90 evolutionary calculation between 0 and 640 function evaluations. We know from the values in Table 2 that the evolutionary algorithm has not yet begun its rapid increase in this region. Because each step of our simulated annealing program needs six function evaluations on average (one for each valid SMILES string produced by a mutation operator) but our evolutionary algorithm needs exactly twenty function evaluations for each generation (one for each child), we need to perform a much longer simulated annealing calculation to provide a more accurate comparison.\par
Because many quantum chemical methods require large amounts of CPU time to determine the properties of a molecule, any optimization algorithm that wants to find molecules with specific properties will want to use as few function evaluations as possible. Our evolutionary calculations suggest that crossover operators play an important role in increasing the speed of convergence in these calculations. The literature contains numerous suggestions on how we could improve this program and in a future paper we plan to investigate several of these ideas \cite{Ref24,Ref25}. Although they started from promising NLO molecules, both the simulated annealing and evolutionary calculations were able to substantially increase the value of $\beta$. The structure of this molecule with the highest value is shown in Figure 3. The obvious next step would be to confirm this value using a more accurate quantum chemical method and/or simultaneously optimize some of the other physical requirements that a successful nonlinear optical material will need. We hope to perform this analysis in the near future.\par

\begin{figure}[h]
  \centering
  \includegraphics[width=\linewidth]{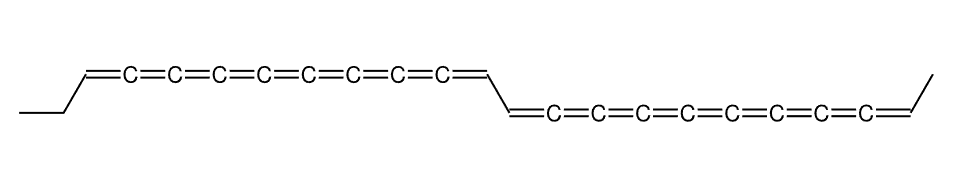}
  \caption{Molecule produced by our 10/90 evolutionary program with $\beta$=649417 atomic units.}
\end{figure}

\begin{acks}
The authors thank Dr. Jacob Schrum for many helpful suggestions.
\end{acks}

\bibliographystyle{ACM-Reference-Format}

\end{document}